\documentclass[twocolumn,aps,floatfix,citeautoscript,prmaterials,superscriptaddress,10pt,cite]{revtex4-2} %

\usepackage{graphicx}

\usepackage{color}
\usepackage{amsmath}
\usepackage{bbold}
\usepackage[caption=false,labelformat=empty,position=top,font=bf]{subfig}

\usepackage{xcolor}
\usepackage[colorlinks=true,linkcolor=blue,urlcolor=blue,citecolor=blue]{hyperref}

\begin{document}

\title{Strain-tunable magnetic and magnonic states in Ni-dihalide monolayers}

\author{Ali Ghojavand}
\affiliation{Department of Physics \& NANOlab Center of Excellence, University of Antwerp, Groenenborgerlaan 171, B-2020 Antwerp, Belgium}

\author{Maarten Soenen}
\affiliation{Department of Physics \& NANOlab Center of Excellence, University of Antwerp, Groenenborgerlaan 171, B-2020 Antwerp, Belgium}

\author{Nafise Rezaei}
\affiliation{Skolkovo Institute of Science and Technology, 121205, Bolshoy Boulevard 30, bld. 1, Moscow, Russia}

\author{Mojtaba Alaei}
\affiliation{Skolkovo Institute of Science and Technology, 121205, Bolshoy Boulevard 30, bld. 1, Moscow, Russia}
\affiliation{Department of Physics, Isfahan University of Technology, Isfahan 84156-83111, Iran}

\author{Cem Sevik}
\affiliation{Department of Physics \& NANOlab Center of Excellence, University of Antwerp, Groenenborgerlaan 171, B-2020 Antwerp, Belgium}

\author{Milorad V. Milo\v{s}evi\'c}
\email{milorad.milosevic@uantwerpen.be}
\affiliation{Department of Physics \& NANOlab Center of Excellence, University of Antwerp, Groenenborgerlaan 171, B-2020 Antwerp, Belgium}
\affiliation{Instituto de Física, Universidade Federal de Mato Grosso, Cuiabá, Mato Grosso 78060-900, Brazil}

\begin{abstract}
Monolayer NiI$_2$ garners large research interest due to its multiferroic behavior stemming from the interplay between its non-collinear magnetic order and the spin-orbit coupling. This prompts an investigation into the stability of the magnetic order in NiI$_2$ and similar materials under external stimuli. In this work, we report the effect of biaxial and uniaxial strain on the magnetic ground state, the critical temperature, and the magnonic properties of the NiX$_2$ (X = I, Br, Cl) monolayers. For all three materials, we reveal intricate strain-dependent phase diagrams, including ferromagnetic, helimagnetic, and skyrmionic phases. Moreover, we discuss the necessity of considering the biquadratic exchange interaction in the latter analysis. We reveal that the biquadratic exchange significantly alters both the magnetic ground state and the critical temperature of the magnetic order, and we demonstrate that its importance becomes even more explicit when monolayer Ni-dihalides are strained. Finally, we calculate the magnonic dispersion for the predicted magnetic states, showing that the skyrmionic phase functions as a magnonic crystal, and demonstrate the presence of strain-tunable soft magnon modes at finite wavevectors in the helimagnetic phase. 
\end{abstract}

\date{\today}

\maketitle

\section{\label{sec:intro}introduction}
Inspired by the successful synthesis of atomically thin CrI$_3$ \cite{huang2017layer}, and Cr$_2$Ge$_2$Te$_6$ \cite{gong2017discovery}, magnetic van der Waals (vdW) materials have attracted a widespread interest in both experimental and theoretical research, providing a platform for the exploration of long-range magnetic order down to the ultimate two-dimensional (2D) limit, and showing promise for the development of spintronic devices and technological applications \cite{elahi2022review,lin2019two,li2019intrinsic, zhang2021gigantic}. 

A notable subset of magnetic vdW materials are the nickel dihalides (NiX$_2$, where X = I, Br, Cl). These insulating 2D magnets are composed of nickel cation layers with partially filled $3d$ electronic shells, paired with non-magnetic halide anions (Cl, Br, and I). The octahedral coordination of nickel ions leads to a division of the $d$-shell electronic states into two distinct sets of orbitals, namely three fully-occupied $t_{2g}$ orbitals and the two partially occupied $e_g$ orbitals, equipping the Ni$^{2+}$ ions with a nominal magnetic moment of 2 $\mu_\mathrm{B}$ \cite{mcguire2017crystal}. 

Bulk NiX$_2$ crystals show various magnetic ground states, including helimagnetism, also known as a spiral (SP) phase, in both NiI$_2$ and NiBr$_2$ and ferromagnetism (FM) in NiCl$_2$ \cite{mcguire2017crystal,adam1980neutron,day1980incommensurate, kuindersma1981magnetic, lindgard1975spin, ni2021giant}. The SP phases in NiI$_2$ and NiBr$_2$ are especially interesting, as they break inversion symmetry, creating a finite polarization, and thus, exhibiting multiferroic behavior \cite{xiang2011general, spaldin2017multiferroics, katsura2005spin, ni2021giant, 2021noncollinear}. Interestingly, recent studies have shown that both the magnetic and the multiferroic properties of bulk NiX$_2$, persist down to the monolayer limit \cite{song2022evidence,ni2021giant,fumega2022microscopic, das2024revival, amini2024atomic}. Furthermore, recent theoretical studies further suggest that the NiX$_2$ monolayers could potentially host chiral magnetic phases such as skyrmions (Sk) and antibiskyrmions (A2Sk), even in the absence of an antisymmetric contribution to the exchange, which is better known as the Dzyaloshinskii-Moriya interaction (DMI) \cite{amoroso2020spontaneous}.

On the other hand, strain engineering has emerged as a powerful technique for tuning the properties of nanomaterials,  employing flexible substrates~\cite{conley2013bandgap,peng2014tuning,dai2019strain,zhao2021piezo} to induce uniaxial strain or the piezo-substrates ~\cite{hui2013exceptional,won2019flexible,ding2010stretchable} to generate biaxial strain. This technique is particularly highlighted in various 2D van der Waals (vdW) magnets such as CrX$_3$ \cite{webster2018strain,zheng2018strain,dupont2021monolayer,menezes2022,soenen2023strain}, CrTe$_3$ \cite{guo2014tuning}, Cr$_2$Ge$_2$Te$_6$ \cite{o2022enhanced}, and Fe$_4$GeTe$_2$ \cite{wang2023interfacial}, where their remarkable mechanical flexibility provides a convenient way to precisely tune magnetic exchange interactions and anisotropy. However, previous theoretical studies on the magnetic behavior of strained NiX$_2$ monolayers \cite{mushtaq2017nix, lu2019mechanical, kulish2017single, han2020enhanced,ni2022plane} have largely employed simple spin Hamiltonian models, such as the Ising model including only nearest-neighboring (NN) exchange interactions. Such models have not only neglected interactions beyond the NN, which are crucial for understanding magnetic frustration in these materials but also overlooked higher-order terms such as the biquadratic (BQ) exchange interaction. Recent studies \cite{ni2021giant,kartsev2020biquadratic} have highlighted the importance of these higher-order terms for a comprehensive understanding of the magnetic properties of 2D magnets, especially those with edge-shared octahedra. 
Therefore, to accurately predict magnetic properties such as the ground-state spin configurations and critical temperature of strained NiX$_2$ monolayers, it is essential to extend the spin hamiltonian model to include all significant interactions. This includes the BQ interaction and contributions due to the spin-orbit coupling (SOC), such as the single-ion anisotropy (SIA), the DMI, and the exchange anisotropy.

In the present work, we have conducted systematic first-principles calculations using Density Functional Theory (DFT) to parameterize a Heisenberg spin model that includes the BQ exchange, and subsequently performed the Monte Carlo (MC) simulations, to study the effect of both biaxial and uniaxial strain on the ground-state spin configuration and the critical temperatures of the NiX$_2$ monolayers. We found that magnetic frustration due to the competing FM NN exchange and the antiferromagnetic (AFM) third-nearest neighbor (3NN) exchange can be significantly tuned by the application of both biaxial and uniaxial strain. Such tuning leads to various magnetic states, including SP, FM, and even A2Sk phases. Further, by comparing strain-dependent phase diagrams for both cases with and without the inclusion of a BQ coupling term in the Hamiltonian, we demonstrate the role played by the BQ exchange in reducing frustration, thereby influencing not only the magnetic ground state but also the critical temperature of the NiX$_2$ monolayers. Finally, we investigate the behavior of magnons in FM, SP, and A2Sk phases, revealing a gapped dispersion for the latter which confirms the magnonic crystal functionality of skyrmion lattices~\cite{ma2015,chen2021review,wang2020}. In addition, we confirm the recent prediction of soft magnon modes in the SP phase of NiI$_2$ \cite{cong2024soft} and predict the same modes for the SP phases of NiBr$_2$ and NiCl$_2$. We show that these modes can be significantly enhanced, induced, or removed by application of biaxial and uniaxial strain. These soft magnon modes are a potentially rich source of novel physics manifestations, as they could host magnon Bose-Einstein condensates or drive magnetic phase transitions at finite temperatures \cite{cong2024soft}. 

This paper is organized as follows. In section \ref{sec:method}, we detail the computational methodology used in this work, including discussions of the Heisenberg spin model, the performed DFT and MC simulations, and the numerical calculations of the magnonic dispersions. Further, in section \ref{sec:magnetic_properties}, we discuss the exchange parameters, the magnetic ground state, and the critical temperature of the pristine NiX$_2$ monolayers in section \ref{sec:pristine}, and discuss how these properties evolve as a function of biaxial and uniaxial strain in sections \ref{sec:biaxial} and \ref{sec:uniaxial} respectively. Finally, in section \ref{sec:magnons}, we discuss the magnonic properties under both biaxial and uniaxial strain (in sections \ref{sec:biaxial_magnons} and \ref{sec:uniaxial_magnons} respectively). Section \ref{sec:conclusion} summarizes our results.

\section{\label{sec:method}methodology}
\begin{figure}[t!]
\includegraphics[width=\linewidth]{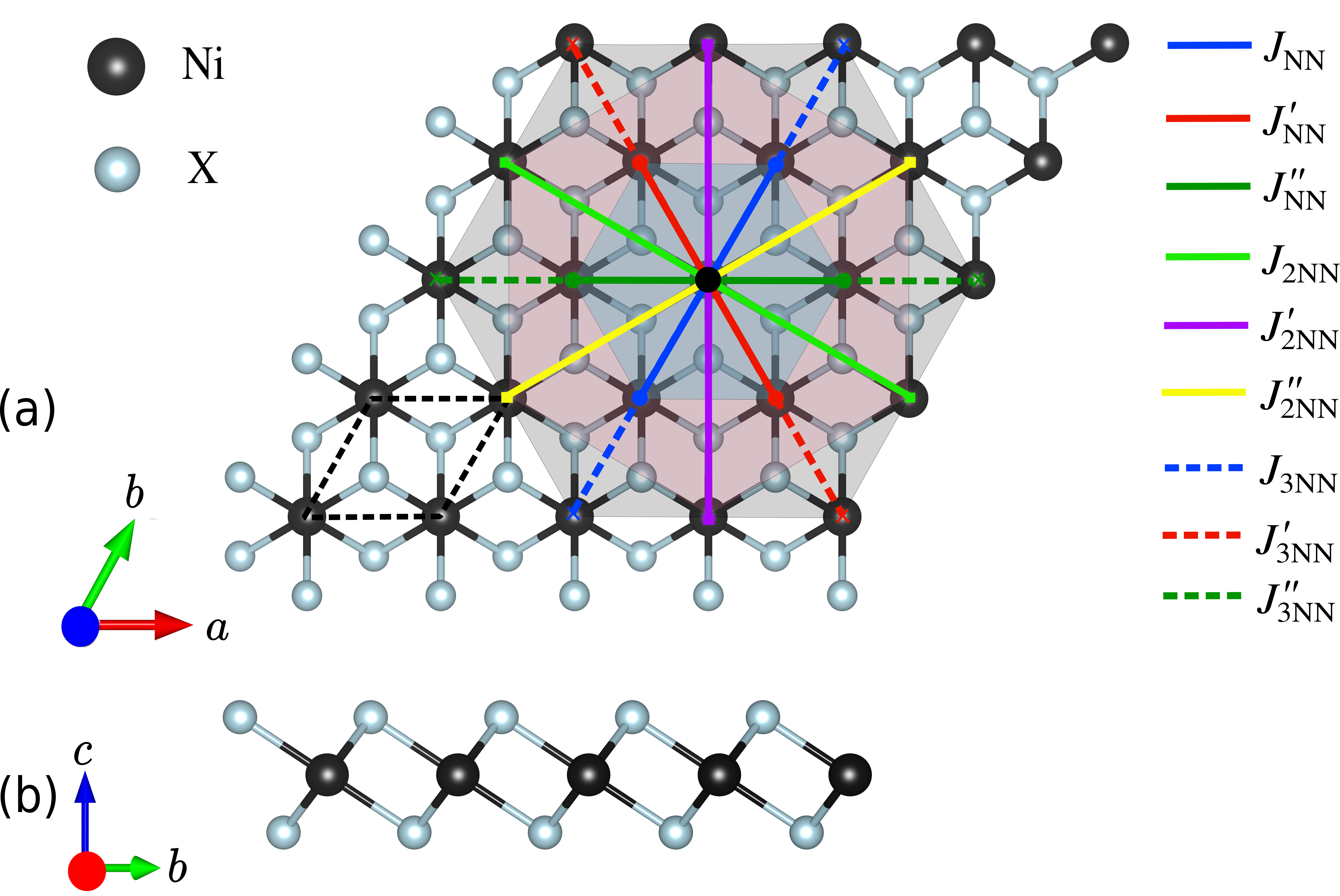}%
\caption{\label{fig:structure} Top view (a) and side view (b) of the crystal structure of a NiX$_2$ monolayer. The nickel and halide atoms are depicted with black and gray spheres, respectively. Corresponding exchange pairs are indicated in the figure. Blue, red, and dark-green solid (dashed) lines correspond to the NN (3NN) exchange, respectively. Light-green, purple, and yellow lines correspond to the 2NN pairs. The unit cell is marked with a black dotted line. The crystal structure was drawn using \textsc{vesta} \cite{VESTA}.(c) Shows a fragment of a triangular lattice of magnetic spins. The solid angle ($\Omega$) corresponds to the area of a spherical triangle formed by the spin vectors $\vec{S_1}$, $\vec{S_2}$, and $\vec{S_3}$ located at the vertices of the triangular plaquette.}
\end{figure}

The magnetic interactions of the NiX$_2$ monolayers are described by the following Heisenberg spin Hamiltonian:
\begin{align}\label{eq:hamiltonian}
\mathcal{H} = \underbrace{\sum_{i} \vec{S}_i \cdot \mathcal{A}_{ii} \cdot \vec{S}_i}_{\text{SIA}} + \underbrace{\sum_{i<j} \vec{S}_i\cdot \mathcal{J}_{ij} \cdot \vec{S}_j}_{\mathcal{H}_{\text{ex}}} + \underbrace{\sum_{i<j}K_{ij}(\vec{S}_i\cdot \vec{S}_j)^2}_{\text{BQ}}, 
\end{align}%
where $\vec{S}_i$ represents the spin unit vector at the ith lattice site, $\mathcal{A}_{ii}$ and $\mathcal{J}_{ij}$ are the SIA and exchange tensors, respectively, and the last term accounts for the BQ exchange interaction. Due to the inherent rotational symmetry of the systems under scrutiny, only the out-of-plane component of the SIA tensor is non-zero ($\mathcal{A}_{ii}^{zz} \equiv A^{zz}\ \forall\ i$). However, when the rotational symmetry is broken, for example, by applying uniaxial strain, the full SIA-tensor needs to be calculated. The exchange part of the Hamiltonian can be decomposed into three contributions as:
\begin{align}\label{eq:exchange}
\mathcal{H}_\mathrm{ex} =  \sum_{i<j}\left[J_{ij} \vec{S}_i \cdot \vec{S}_j + \vec{D}_{ij} \cdot (\vec{S}_i \times \vec{S}_j) + \vec{S}_i\cdot \Gamma_{ij} \cdot \vec{S}_j \right], 
\end{align}
where the first term is the isotropic exchange coupling (\( J_{ij} = \frac{1}{3} \mathrm{Tr} \mathcal{J}_{ij} \)), the second term represents the DMI, which is calculated as \(\vec{D}_{ij} = \frac{1}{2} (\mathcal{J}_{ij}- \mathcal{J}_{ij}^\mathrm{T})\) but vanishes in all the systems under investigation in this work due to the presence of inversion symmetry between all considered Ni-Ni pairs, and the last term accounts for the anisotropic part of the exchange tensor \(\Gamma_{ij} = \frac{1}{2} (\mathcal{J}_{ij} + \mathcal{J}_{ij}^\mathrm{T}) - J_{ij} \mathbb{1} \). Note that, for the exchange tensor, we consider interactions up to 3NN pairs as exemplified in Figure~\ref{fig:structure}(a). For the BQ exchange, we only include the NN interactions, which has been shown to be sufficient for the systems under scrutiny \cite{fedorova2015biquadratic, wysocki2011consistent, turner2009kinetic, harris1963biquadratic, ni2021giant, kartsev2020biquadratic}. 

In order to determine the magnetic parameters specified in Eq. (\ref{eq:hamiltonian}), we utilized the four-state energy mapping (4SM) methodology, which involves selecting four distinct spin configurations for each exchange parameter of every spin pair and mapping the DFT energies of these configurations to the corresponding Heisenberg Hamiltonians \cite{sabani2020ab,xiang2013magnetic}. For both the biaxial and the uniaxial cases, we applied strain values ranging from -8\% compressive strain to 8\% tensile strain. For each strain value, the atomic positions were fully relaxed. In the context of biaxial strain, which is uniformly applied along both the $a$- and $b$-axes, the system’s inherent characteristics, such as rotational symmetry and bond equivalency, are conserved. Hence, in all systems under biaxial strain, it suffices to calculate only one exchange tensor, $\mathcal{J}_{ij}$ and derive the other exchange parameters, i.e. $\mathcal{J}_{ij}' = R_z(\frac{2\pi}{3})\cdot \mathcal{J}_{ij}\cdot R_z^T(\frac{2\pi}{3})$ and $\mathcal{J}_{ij}'' = R_z(\frac{2\pi}{3})\cdot \mathcal{J}_{ij}'\cdot R_z^T(\frac{2\pi}{3})$ (see Figure~\ref{fig:structure}(a)), using the rotation matrix around the out-of-plane axis $R_z$.  The uniaxial strain in this study is applied in such a way that strain is only applied along the $b$-direction while the $a$-direction remains unchanged. This application of strain breaks the rotational symmetry and results in changes in bond equivalence. As a result, all exchange pairs must be calculated independently. The same argument holds for the BQ interaction, yielding only one parameter in the systems under biaxial strain and three unique parameters when uniaxial strain is applied.

For all DFT calculations, we used the projector-augmented wave (PAW) method \cite{bl1994hl} as implemented in the Vienna Ab initio Simulation Package (VASP) \cite{kresse1999ultrasoft, kresse1993ab, kresse1996efficiency}. We employ the Perdew-Burke-Ernzerhof (PBE) \cite{perdew1996generalized} exchange-correlation functional within the generalized gradient approximation (GGA). A plane-wave cutoff of 550 eV was used for all the studied cases. We selected sufficiently large supercells (6$\times$6$\times$1) to ensure accurate results and to avoid periodic interactions between Ni-atoms and their periodic images. The Brillouin zone (BZ) was sampled by a 12$\times$12$\times$1 Monkhorst-Pack $k$-point mesh for simulations on a unit cell and a  2$\times$2$\times$1 $k$-point mesh when using supercells. To account for the periodic boundary conditions in the out-of-plane direction, we include 15 \AA{} of vacuum in the unit cell. The effective Hubbard parameter U\textsubscript{eff} (U\textsubscript{eff}= U – J) is imposed on the $3d$ orbitals of the Ni-ions using Dudarev's approach \cite{liechtenstein1995density, dudarev1997effect}. To estimate U\textsubscript{eff}, we adopted the linear response method~\cite{cococcioni2005linear}, yielding values of 4.84 eV, 4.59 eV, and 4.47 eV for NiI\textsubscript{2}, NiBr\textsubscript{2}, and NiCl\textsubscript{2}, respectively, which are in good agreement with earlier studies \cite{ni2021giant, mcguire2017crystal}. All structures were optimized using DFT calculations with FM spin ordering until the energy differences between consecutive ionic steps were less than $10^{-6}$ eV/supercell. For the 4SM calculations, we performed non-collinear DFT calculations with the inclusion of SOC in the constrained local moment approach.

In order to evaluate the dynamical stability of the strained structures, we computed the force constants of ($6\times6\times1$) supercells in real space using density functional perturbation theory (DFPT) \cite{baroni2001phonons} as implemented in the PHONOPY code~\cite{togo2015first}. The phonon dispersion curves, which confirm the stability of the structures under investigation, are presented in section SIII of the Supplementary Material~\cite{supplm}. 

The critical temperature and spin configurations were obtained through parallel tempering Monte Carlo (PTMC)~\cite{swendsen1986replica} simulations, utilizing the Esfahan spin simulation package (ESpinS)~\cite{rezaei2022espins}, an open-source classical spin MC software. MC simulations were carried out on $40\times 40 \times 1$ supercells with periodic boundary conditions. We used $10^{6}$ MC steps per spin for thermal equilibration and data collection. In the PTMC algorithm, we allowed the spin configurations at different temperatures to swap after each 40 MC step. 

The topological nature of the obtained magnetic structures is assessed by computation of the topological charge. This charge is determined using the discrete lattice model~\cite{BERG1981412}, where the topological contribution from each triangular plaquette is given by the solid angle \( \Omega \) subtended by three neighboring spins. The solid angle is determined using the following expression\cite{amoroso2020spontaneous}:
\begin{align}\label{eq:topch}
\tan\left(\frac{\Omega}{2}\right) = \frac{\vec{S_1} \cdot \vec{S_2} \times \vec{S_3}}{1 + \vec{S_1} \cdot \vec{S_2} + \vec{S_2} \cdot \vec{S_3} + \vec{S_1} \cdot \vec{S_3}}.
\end{align}%
where $\vec{S_1}$,  $\vec{S_2}$, and  $\vec{S_3}$ are the unit vectors representing the spin directions at the vertices of a triangular plaquette (see Fig.\ref{fig:structure}(c)). The total topological charge is then obtained by summing over all plaquettes within the magnetic unit cell.

Additionally, to differentiate between magnetic phases, we computed the spin structure factor as:
\begin{equation}\label{eq:structurefact}
S(\vec{q}) = \frac{1}{N} \sum_\alpha \left\langle \left| \sum_{i} S_{i}^{\alpha} e^{-i \vec{q} \cdot \vec{r}_i} \right|^2 \right\rangle,
\end{equation}
where $\vec{r}_i$ is the position of spin $\vec{S}_i$, $N$ is the total number of spins in the MC simulation cell, and the summation runs over the spin components $\alpha= \{x, y, z\}$.

\begin{table}[b!]
\caption{\label{tab:params} Magnetic parameters for the pristine NiX$_2$ monolayers including the isotropic NN ($J_\mathrm{NN}$), 2NN ($J_\mathrm{2NN}$), and 3NN ($J_\mathrm{3NN}$) exchange parameters, the NN anisotropic exchange parameters \(\Gamma_\mathrm{NN}^{xx}\), \(\Gamma_\mathrm{NN}^{yy}\), \(\Gamma_\mathrm{NN}^{zz}\) and \(\Gamma_\mathrm{NN}^{yz}\), the SIA parameter for each spin site $A^{zz}$, and the NN-BQ exchange parameter $K_\mathrm{NN}$. Further, several important ratios as well as the critical temperatures for both with and without the inclusion of the BQ exchange are presented. The magnetic parameters and temperatures are expressed in units of meV and K, respectively.}
\begin{ruledtabular}
\begin{tabular}{c c c c} 
  & NiI$_2$ & NiBr$_2$ & NiCl$_2$ \\ 
 \hline
 $J_\mathrm{NN}$ & -3.49 & -2.97 & -2.60 \\
 $J_\mathrm{2NN}$ & -0.15 & -0.09 & -0.05 \\
 $J_\mathrm{3NN}$ & 2.50 & 1.28 & 0.77 \\
 \hline
 \(\Gamma_\mathrm{NN}^{xx}\) & -0.42 & -0.04 & 0.00 \\
 \(\Gamma_\mathrm{NN}^{yy}\) & 0.52 & 0.04 & 0.00 \\
 \(\Gamma_\mathrm{NN}^{zz}\) & -0.09 & 0.00 & 0.00 \\
 \(\Gamma_\mathrm{NN}^{yz}\) & -0.59 & -0.05 & 0.00 \\
 \hline
 $A^{zz}$ & 0.07 & 0.02 & 0.02 \\
 \hline
 $K_\mathrm{NN}$ & -0.67 & -0.64 & -0.66 \\ 
  \hline
 $|J_\mathrm{3NN}/J_\mathrm{NN}|$ & 0.72 & 0.43 & 0.30 \\
 \(\Gamma_\mathrm{NN}^{yz}/J_\mathrm{NN}\) & 0.17 & 0.02 & 0.00 \\
 $K_\mathrm{NN}/J_\mathrm{NN}$ & 0.20 & 0.22 & 0.25 \\
 \hline
 $T_c^{K_\mathrm{NN} \neq 0}$ & 6.60 & 4.10 & 15.50 \\ 
 $T_c^{K_\mathrm{NN} = 0}$ & 11.60  & 5.20  & 4.00 \\ 
\end{tabular}
\end{ruledtabular}
\end{table}

We investigate the magnonic properties of the different magnetic phases using \textsc{spinw} \cite{spinw}, a code based on linear spin-wave theory that includes an algorithm to calculate the magnonic dispersion for single-$\mathbf{q}$ incommensurate magnetic structures, which can therefore be conveniently applied to the SP and FM phases under scrutiny in this work. Further, the method can be extended to multi-$\mathbf{q}$ magnetic structures by constructing $\mathbf{q} = 0$ magnetic supercells \cite{spinw}, however, for incommensurate multi-$\mathbf{q}$ structures, like the A2SK phase discussed in this work, this yields an approximate solution. Additionally, note that, \textsc{spinw} does not allow for inclusion of the BQ exchange for magnetic structures with $\mathbf{q} \neq 0$. The main effect of the BQ term is that it can result, for some strain values, in a different magnetic ground state (see Sec. \ref{sec:magnetic_properties}), which will evidently result in different magnonic behavior. Further, the BQ exchange can change the energies of the magnon modes that will be discussed in Sec. \ref{sec:magnons}, as has also been shown by Cong and Shen \cite{cong2024soft}, however, we are confident that the strain dependence of these modes will be qualitatively similar even without inclusion of the BQ term.

\section{\label{sec:result}Results}
\subsection{\label{sec:magnetic_properties}Magnetic properties}
\subsubsection{\label{sec:pristine}Pristine NiX$_2$ monolayers} 
The crystal structure of a NiX$_2$ monolayer is depicted in Figure \ref{fig:structure}. These materials are characterized by the $P\overline{3}m1$ space group, which has a sandwiched (X-Ni-X) structure where the halide ions are located in a close-packed structure, and the nickel atoms occupy the octahedrally coordinated interstitial positions. DFT calculations yield relaxed in-plane lattice constants for NiI$_2$, NiBr$_2$, and NiCl$_2$ equal to 3.97 \AA, 3.69 \AA, and 3.47 \AA, respectively, which are in agreement with previous studies~\cite{amoroso2020spontaneous, ni2021giant, ni2022plane, mcguire2017crystal}. 

\begin{figure*}
\includegraphics[width=\linewidth]{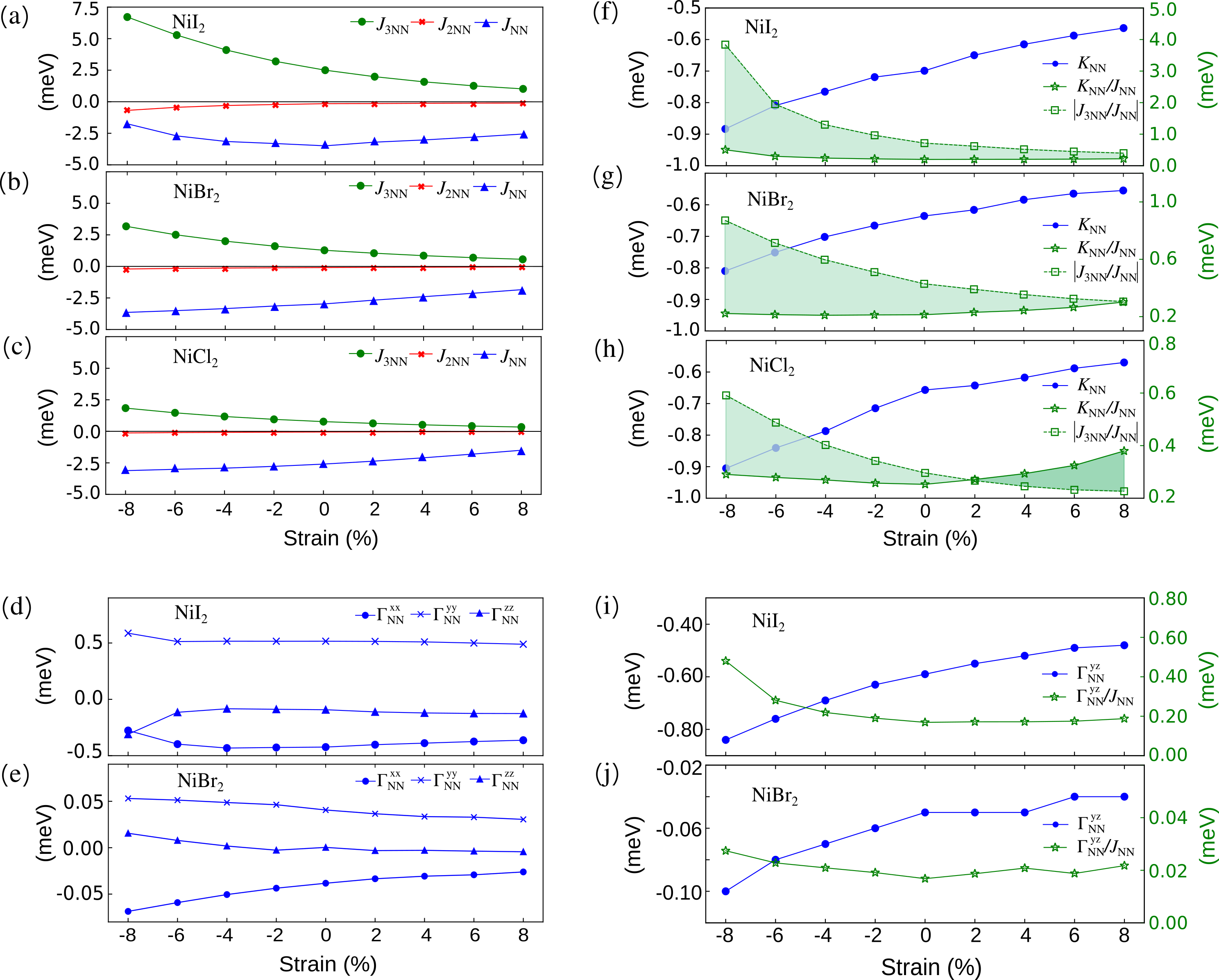}%
\caption{\label{fig:biaxpa}Magnetic parameters of monolayers NiI$_2$, NiBr$_2$, and NiCl$_2$ as a function of biaxial strain. (a-c) NN, 2NN, and 3NN isotropic exchange interactions as a function of biaxial strain for (a) NiI$_2$, (b) NiBr$_2$, and (c) NiCl$_2$, respectively. (f-h) Strain-dependence of the NN-BQ exchange interaction for (f) NiI$_2$, (g) NiBr$_2$, and (h) NiCl$_2$, respectively. (d,e,i,j) Strain dependence of the NN anisotropic exchange parameters. The shaded green regions in (f,g,h) highlight the difference between the $|J_\mathrm{3NN}/J_\mathrm{NN}|$ and $K_\mathrm{NN}/J_\mathrm{NN}$ ratios. The values of those ratios are plotted with respect to the axis on the right, shown in green.}
\end{figure*}
\begin{figure}[t!]
\includegraphics[width=\linewidth]{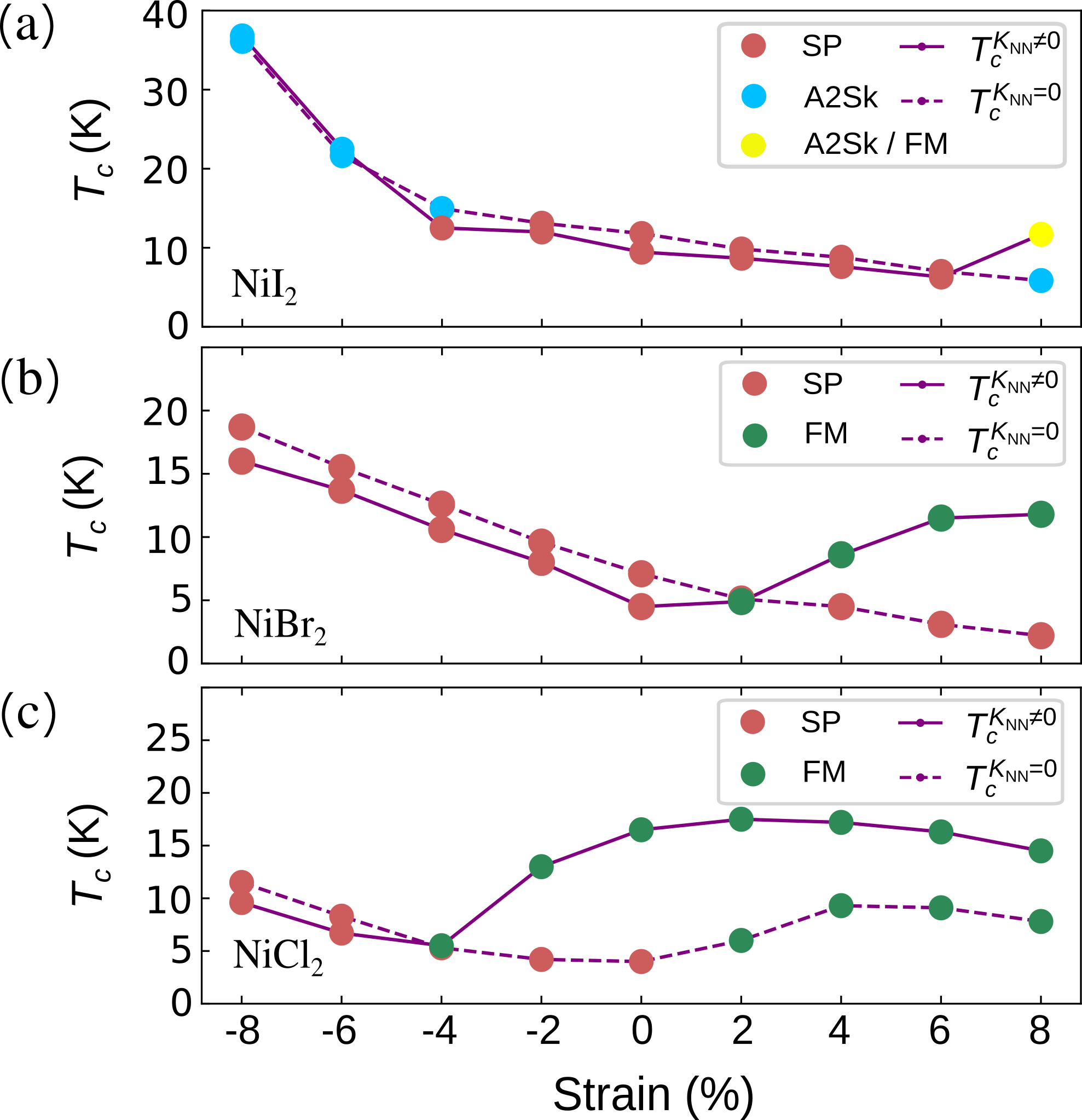}%
\caption{\label{fig:biaxtc} The critical temperatures as a function of biaxial strain for monolayers of NiI$_2$ (a), NiBr$_2$ (b), and NiCl$_2$ (c) as computed by parallel tempering Monte Carlo simulations. Solid and dashed lines depict critical temperatures computed from the Heisenberg Hamiltonian, with ($T_c^{K_\mathrm{NN} \neq 0}$) and without ($T_c^{K_\mathrm{NN} = 0}$) consideration of BQ exchange interactions. The color-coded dots indicate the ground-state spin order corresponding to each strain: green dots represent the FM order, blue dots show the A2Sk order and red dots signify the SP order.}
\end{figure}
In Table~\ref{tab:params}, we report the magnetic parameters obtained by the 4SM method. Negative (positive) values of the exchange parameters refer to FM (AFM) magnetic interaction, while positive (negative) values of $A^{zz}$ indicate easy-plane (easy-axis) anisotropy. 

The main contributors to the magnetic behavior of these systems are the NN exchange ($J_\mathrm{NN}$) and 3NN exchange ($J_\mathrm{3NN}$), which are FM and AFM interactions, respectively. $J_\mathrm{3NN}$ is comparable to $J_\mathrm{NN}$ as it involves $e_g$-$e_g$ hoppings, as explained in detail in Ref. \cite{riedl2022microscopic}. The relatively small (or negligible) value of $J_\mathrm{2NN}$ compared to $J_\mathrm{NN}$ and $J_\mathrm{3NN}$ arises from the $t_{2g}$-$e_g$ hopping, which results in a weak ferromagnetic interaction \cite{riedl2022microscopic}.
In agreement with prior investigations~\cite{ni2021giant, amoroso2020spontaneous}, competition between  $J_\mathrm{NN}$ and $J_\mathrm{3NN}$ terms, characterized by the frustration ratio $|J_\mathrm{3NN}/J_\mathrm{NN}|$, exhibits a growing trend with higher atomic number of the halide, attributed to the expansive ligand $p$ orbitals facilitating the superexchange mechanism~\cite{anderson1959new}. Similarly, there is a trend of increasing exchange anisotropy with stronger ligand SOC along the halide series~\cite{amoroso2020spontaneous}. NiCl$_2$ displays full isotropic magnetic behavior, while NiBr$_2$ is mostly isotropic with minor deviations (\(\Gamma_\mathrm{NN}^{xx}\) $\approx$ \(\Gamma_\mathrm{NN}^{yy}\) $\approx$ \(\Gamma_\mathrm{NN}^{zz}\), and \(\Gamma_\mathrm{NN}^{zy}\) $\approx$ \(\Gamma_\mathrm{NN}^{yz}\) $\approx 0$). The minor anisotropy observed in NiBr$_2$ is essentially negligible (it is almost two orders of magnitude smaller than its isotropic parameters). In contrast, NiI$_2$ demonstrates significant anisotropic behavior, where the anisotropic parameters associated with the nearest-neighbor exchange interaction are one order of magnitude larger than in NiBr$_2$. The ratio \(\Gamma_\mathrm{yz}/J_\mathrm{NN}\), indicating the canting of the anisotropy axes from the direction perpendicular to the monolayers, increases with the ligand variation (from 0.00 for NiCl$_2$ to 0.17 for NiI$_2$). This anisotropic contribution becomes negligible in the interactions of the second and third nearest neighbors. Given the relatively small magnitude of the SIA in comparison to other interactions, its role in stabilizing the magnetic ground states is negligible.

\begin{figure*}
\includegraphics[width=\linewidth]{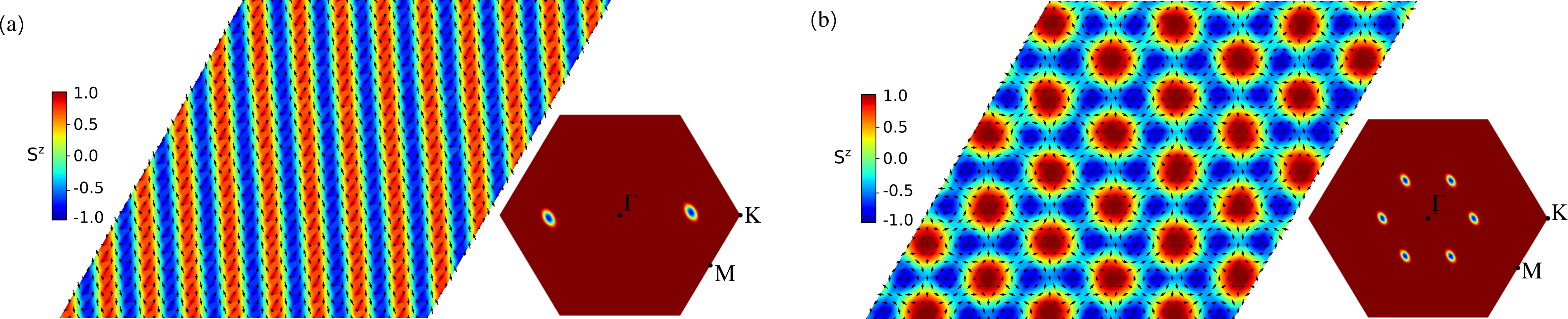}%
\caption{\label{fig:sptext} Low-energy magnetic structures computed from the Heisenberg Hamiltonian with consideration of BQ exchange interactions on a \(40 \times 40 \times 1\) supercell, computed from PTMC simulations. (a) SP phase observed in a NiI$_2$ monolayer under a 2\% tensile biaxial strain. (b) A2Sk phase observed in a NiI$_2$ monolayer under a 6\% compressive biaxial strain. The color map indicates the out-of-plane spin component $S^z$, the in-plane spin components are denoted by black arrows. The spin structure factor ${S}(\vec{q})$ associated with each structure is plotted within the first BZ.}
\end{figure*}
The obtained NN-BQ exchange parameters ($K_\mathrm{NN}$) exhibit similar magnitudes for NiI$_2$, NiBr$_2$, and NiCl$_2$ and have the same sign as $J_\mathrm{NN}$, meaning that the BQ term is promoting a collinear spin orientation and reinforces the FM alignment of the spins. Upon comparison of the $K_\mathrm{NN}/J_\mathrm{NN}$ and $|J_\mathrm{3NN}/J_\mathrm{NN}|$ ratios, it becomes evident that for NiCl$_2$, these ratios are nearly identical, contrasting the NiBr$_2$ and NiI$_2$ cases where the $K_\mathrm{NN}/J_\mathrm{NN}$ ratio is approximately 2 and 3 times smaller than the $|J_\mathrm{3NN}/J_\mathrm{NN}|$ ratio, respectively.  This comparison suggests that in NiCl$_2$, the BQ exchange is strong enough to mitigate or even overcome magnetic frustration. In contrast, for NiBr$_2$ and NiI$_2$, the BQ interaction is insufficient to counteract the frustration, resulting in a less significant impact on spin alignment. 
In order to investigate the influence of NN-BQ interaction on the ground-state spin configurations and the critical temperature, we performed PTMC simulations with and without considering the NN-BQ exchange interaction within the Hamiltonian of Eq. (\ref{eq:hamiltonian}). Consistent with previous studies \cite{ni2021giant}, it is found that for NiCl\textsubscript{2} monolayer, the NN-BQ term overcomes the frustration and stabilizes FM order. Consequently, this results in a substantial difference in critical temperatures predicted for each case, namely $T_c^{K_\mathrm{NN}\neq 0} = 15.5$ K and $T_c^{K_\mathrm{NN}= 0} = 4.0$ K. In contrast, in NiI\textsubscript{2} and NiBr\textsubscript{2} systems, the NN-BQ term is not large enough to overcome frustration, and thus their ground-state spin configurations remain SP.

\subsubsection{\label{sec:biaxial}Effect of biaxial strain}

As discussed in the previous section, the magnetic properties of the pristine NiX$_2$ monolayers mainly depend on three factors, namely, the magnetic frustration ratio ($|J_\mathrm{3NN}/J_\mathrm{NN}|$), the relative size of the BQ exchange $K_\mathrm{NN}/J_\mathrm{NN}$, which hold the capacity to impact or potentially mitigate this state of frustration, and the anisotropic NN exchange $\Gamma_\mathrm{NN}$ which determines the preferential orientation direction of the spins. In this section, we investigate the effects of applying biaxial strain on NiX$_2$ (X = I, Br, and Cl) monolayers to gain insight into how biaxial strain influences the magnetic ordering and critical temperatures.

Before delving into the effects of applying biaxial strain on the magnetic properties of strained systems, we initially assessed the dynamical stability of the considered crystals with 8\% tensile and compressive biaxial strains in their FM state via full phonon spectrum analysis. The dispersions without imaginary frequency through the whole momentum space prove the dynamical stability of these crystals under biaxial deformations, as shown in Fig. S2 in the Supplementary Material~\cite{supplm}. The minor imaginary frequencies observed near the $\Gamma$ point are most likely due to numerical errors or limited size
of supercells used in phonon calculation.

With the application of biaxial strain (see Figure 2(a-c)), the dominant isotropic exchange interactions maintain their signs: $J_\mathrm{NN}$ remains FM, while $J_\mathrm{3NN}$ remains AFM across all strain values. $J_\mathrm{2NN}$ parameter remains small (negligible) in comparison to both $J_\mathrm{NN}$ and $J_\mathrm{3NN}$. The weaker sensitivity of \( J_{\text{NN}} \) to biaxial strain compared to \( J_{\text{3NN}} \) can be attributed to the intricate interplay between superexchange and direct exchange mechanisms among nearest neighbors. Due to the shorter atomic distances, these mechanisms can effectively counterbalance each other, leading to a reduced response of \( J_{\text{NN}} \) to biaxial strain \cite{riedl2022microscopic, fazekas1999lecture}. In contrast, for third nearest neighbors, the magnetic interactions primarily arise from super-superexchange processes, largely driven by $e_g$-$p$ hybridizations~\cite{riedl2022microscopic}, which explains its monotonic response to biaxial strain.

In Figure~\ref{fig:biaxpa}(f-h), one can see the influence of biaxial strain on the NN-BQ exchange interaction ($K_\mathrm{NN}$). In all our systems, the $K_\mathrm{NN}$ parameter exhibits low sensitivity to applied biaxial strain, maintaining values within the range of -0.5 meV to -0.9 meV.  
In order to gain insight into the impact of NN-BQ interaction on the frustration arising from the interplay between $J_\mathrm{NN}$ and $J_\mathrm{3NN}$ parameters, we conducted a comparative analysis of both $|J_\mathrm{3NN}/J_\mathrm{NN}|$ and $K_\mathrm{NN}/J_\mathrm{NN}$ ratios. As illustrated in the shaded area of Figure~\ref{fig:biaxpa}(f-h), the compressive biaxial strain increases the value of $|J_\mathrm{3NN}/J_\mathrm{NN}|$ for all studied monolayers, thereby amplifying the difference between $|J_\mathrm{3NN}/J_\mathrm{NN}|$ and $K_\mathrm{NN}/J_\mathrm{NN}$ ratios. This serves as evidence that compressive strain, by enhancing the effect of magnetic frustration, can intensify the tendency for non-collinear spin ordering, such as SP and A2Sk. This effect is particularly notable in NiI$_2$, reaching the highest magnetic frustration ratio of 3.84 at -8\% compressive strain. On the other hand, applying tensile biaxial strain in all examined systems leads to a decrease in the $|J_\mathrm{3NN}/J_\mathrm{NN}|$ ratio, making it comparable to the $K_\mathrm{NN}/J_\mathrm{NN}$ ratio. In such an alteration, the influence of NN-BQ is highlighted, as it can completely overcome the frustration and promote the FM order. This is particularly evident in the NiCl\textsubscript{2} monolayer, where at 2\% tensile strain, the ratio of $K_\mathrm{NN}/J_\mathrm{NN}$ surpasses $J_\mathrm{3NN}/J_\mathrm{NN}$, indicating a strong tendency towards the formation of FM order.

In Figure~\ref{fig:biaxpa}~(d, e, i, j), we show the impact of biaxial strain on the NN anisotropic exchange parameters, namely, \(\Gamma_\mathrm{NN}^{xx}\), \(\Gamma_\mathrm{NN}^{yz}\), \(\Gamma_\mathrm{NN}^{zz}\), and \(\Gamma_\mathrm{NN}^{yz}\). 
As discussed in section~\ref{sec:pristine}, in pristine NiX$_2$ monolayers, the exchange anisotropy scales with the SOC, which is larger for heavier halide ligands. Indeed, NiCl$_2$ exhibits fully isotropic exchange, while NiBr$_2$ shows nearly isotropic behavior with only minor (negligible) deviations, and NiI$_2$ demonstrates significant anisotropic behavior. This trend persists under biaxial strain: NiCl$_2$ continues to exhibit full isotropic exchange, NiBr$_2$ remains nearly isotropic, and NiI$_2$ shows considerable tunability in its anisotropic exchange parameters. Under biaxial strain, NiI$_2$ achieves comparable magnitudes for the  \(\Gamma_\mathrm{yz}/J_\mathrm{NN}\) and \(|J_\mathrm{3NN}/J_\mathrm{NN}|\) ratios.

\begin{figure*}
\includegraphics[width=\linewidth]{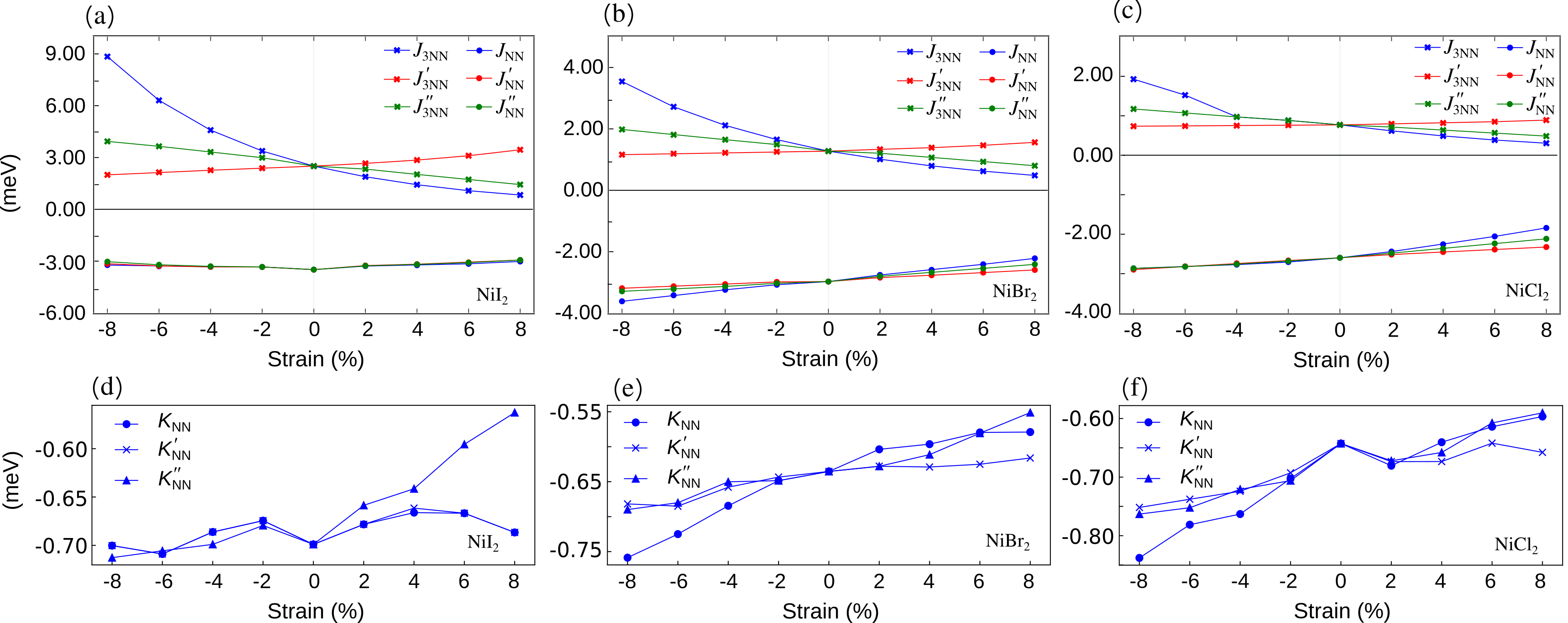}%
\caption{\label{fig:uniaxpa} Magnetic parameters of monolayers NiI$_2$, NiBr$_2$, and NiCl$_2$ as a function of uniaxial strain. The isotropic exchange interactions for the first ($J_\mathrm{NN}$, $J_\mathrm{NN}'$, $J_\mathrm{NN}''$) and third ($J_\mathrm{3NN}$, $J_\mathrm{3NN}'$, $J_\mathrm{3NN}''$) nearest-neighbor interactions are depicted for (a) NiI$_2$, (b) NiBr$_2$, and (c) NiCl$_2$ monolayers. The NN-BQ exchange interactions ($K_\mathrm{NN}$, $K_\mathrm{NN}'$, $K_\mathrm{NN}''$) are shown for (d) NiI$_2$, (e) NiBr$_2$, and (f) NiCl$_2$ monolayers. For better visibility, only the first and third nearest neighbor exchange parameters are depicted. The exchange parameters associated with the second-nearest neighbor are negligible (close to zero) and therefore omitted in this figure. Detailed tables containing all calculated exchange tensors, and SIA parameters are provided in the Supplementary Material~\cite{supplm}.}
\end{figure*}

In order to investigate the effects of biaxial strain on magnetic ground states and critical temperatures, and to gain a deeper understanding of the impact of magnetic parameters, particularly the NN-BQ exchange parameters, on these states, we conducted PTMC simulations using Eq. (\ref{eq:hamiltonian}) across all considered strain values. 
The results are presented in Figure~\ref{fig:biaxtc} under two conditions:  $(i)$ when $K_{\mathrm{NN}} = 0$, with the corresponding critical temperature denoted as $T_c^{K_\mathrm{NN} = 0}$, and  $(ii)$ when $K_\mathrm{NN} \neq 0$,  with the corresponding critical temperature denoted as $T_c^{K_\mathrm{NN} \neq 0}$. In NiCl$_2$, the $T_c^{K_\mathrm{NN} \neq 0}$ shows a nonmonotonic dependence on the compressive strain. It initially decreases until it reaches a compressive strain of -4\%, at which point it starts to increase. Interestingly, this change in the trend of $T_c^{K_\mathrm{NN} \neq 0}$ coincides with a change in the ground-state magnetic order, transitioning from FM to SP order. This can be attributed to the increasing influence of magnetic frustration under compressive strain, given that the value of the $|J_\mathrm{3NN}/J_\mathrm{NN}|$ ratio becomes nearly 1.5 times greater than the $K_\mathrm{NN}/J_\mathrm{NN}$ ratio. On the other hand, the application of tensile biaxial strain on the NiCl$_2$ monolayer does not significantly alter the critical temperature and maintains the FM order as the ground-state. A phase transition from a SP phase to a FM phase is observed under tensile strain of 2\% with $K_\mathrm{NN} = 0$. This phase transition is directly linked to the reduction in the frustration ratio due to the decreasing trend of $J_\mathrm{3NN}$ under tensile biaxial strain. 
The differences observed between $T_c^{K_\mathrm{NN} \neq 0}$ and $T_c^{K_\mathrm{NN} = 0}$ in NiCl$_2$ with tensile strains greater than 2\%, despite observing FM order as the ground-state, can be attributed to the impact of NN-BQ exchange interaction. This finding is similar to observations in other studied 2D systems, where the inclusion of the NN-BQ exchange interaction enhances magnetization, subsequently leading to an increase in the critical temperature \cite{kartsev2020biquadratic}. 

In NiBr$_2$ (see Figure~\ref{fig:biaxtc}(b)), $T_c^{K_\mathrm{NN} = 0}$ displays a linear trend under both compressive and tensile biaxial strain, with no observed phase transition, maintaining the ground-state SP magnetic order. However,  $T_c^{K_\mathrm{NN} \neq 0}$ under compressive strain exhibits a similar increasing trend without any phase transition. Interestingly, under tensile strain, an increment in $T_c^{K_\mathrm{NN} \neq 0}$ coincides with a notable phase transition from SP to FM order.  

As depicted in Figure~\ref{fig:biaxtc}(a), in NiI$_2$, the alterations in $T_c^{K_\mathrm{NN} \neq 0}$ and $T_c^{K_\mathrm{NN} = 0}$ under both compressive and tensile strains are generally similar. However, an exception is observed at 8\% tensile strain, where $T_c^{K_\mathrm{NN} \neq 0}$ experiences a sudden increase. This increase occurs concurrently with the change in the phase diagram to a quasi-degenerate A2Sk and FM phases. We conducted a comparison of the MC energies for the A2Sk and FM spin textures that were obtained in NiI$_2$ at 8\% tensile strain. This comparison revealed a minor energy difference of about 0.3 meV per spin between the A2Sk and FM states, with the FM state being energetically favorable. To further investigate the energy levels of these spin textures, we performed DFT calculations. Specifically, we extracted the spin textures from the MC simulations and constrained the magnetic moment directions according to these textures. We then carried out DFT calculations to assess the energy differences from a DFT perspective. This approach allowed us to verify the MC simulation results with a more precise energy calculation method. However, our DFT calculations showed that the A2Sk order has a slightly lower energy, by about 0.03 meV per spin. This difference is within the error bar of the DFT estimates, suggesting that these two states are nearly energetically equivalent.

NiI$_2$ also shows another phase transition to A2Sk with twice the topological charge of skyrmions ($|Q| = 2$) under compressive strain. Remarkably, this transition occurs regardless of whether the NN-BQ exchange interaction is taken into account. As discussed in Ref.~\cite{amoroso2020spontaneous}, the anisotropic exchange $\Gamma_\mathrm{NN}^{yz}$ serves as a direct measure of the expected deviation from collinearity in the spin arrangement and can play a key role in stabilizing noncollinear spin orders, such as the A2Sk spin order. Indeed, the emergence of this noncollinear spin order results from the competition between the isotropic exchange interaction $J_\mathrm{NN}$ and the anisotropic term $\Gamma_\mathrm{NN}^{yz}$, which can be quantified through the ratio $\Gamma_\mathrm{NN}^{yz}/J_\mathrm{NN}$. As previously mentioned, biaxial strain modifies this ratio in NiI$_2$. Our results indicate that when the $\Gamma_\mathrm{NN}^{yz}/J_\mathrm{NN}$ ratio exceeds approximately 0.2, the system demonstrates a clear tendency to stabilize the A2Sk spin order. A thorough analysis was conducted to examine the contributions of the $\Gamma_\mathrm{NN}^{yz}$ term to the formation of the A2Sk phase in the NiI$_2$ monolayer under -8\% strain. Detailed findings from this analysis can be found in section SV of the supplementary material~\cite{supplm}. In Figure~\ref{fig:sptext}, an illustration of two important discovered magnetic spin configurations along with their corresponding spin structure factors is presented.

\subsubsection{\label{sec:uniaxial}Effect of uniaxial strain}

To further investigate the effects of various strains on the magnetic properties of NiX$_2$ (X = I, Br, Cl) monolayers, we examine the case of uniaxial strain. 
Similar to our approach with biaxial strain, our initial step involves assessing the dynamical stability of the considered crystals with 8\% tensile and compressive uniaxial strains in their FM state via full phonon spectrum analysis. Results in Fig. S3 (Supplementary Material~\cite{supplm}) show that the phonon dispersions are positive over the whole Brillouin zone, except for an extremely small and even negligible imaginary frequency at the $\Gamma$ point because it is within the numerical noise.

Figures~\ref{fig:uniaxpa}(a-c) illustrate the evolution of the magnetic isotropic exchange interactions with the uniaxial strain level. The 1NN and 2NN isotropic exchange parameters show minimal sensitivity to the uniaxial strain, a characteristic also observed under biaxial strain conditions. On the other hand, the 3NN isotropic exchange parameters ($J_\mathrm{3NN}$, $J_\mathrm{3NN}'$, $J_\mathrm{3NN}''$) exhibit remarkable sensitivity to uniaxial strains, with each pair exhibiting distinct behaviors.

As mentioned in section \ref{sec:method}, applying uniaxial strain alters bond equivalency. Consequently, the $J_\mathrm{3NN}$ parameter can be finely tuned through uniaxial strain because of its high sensitivity to variations in bond length. Conversely, the $J_\mathrm{3NN}'$ and $J_\mathrm{3NN}''$ parameters demonstrate insignificant linear variation in response to uniaxial strain. Under 8\% compressive strain, $J_\mathrm{3NN}$ in all systems becomes more than three times larger than $J_\mathrm{3NN}'$. The tensile strain appears to promote FM tendencies, as evidenced by the decreasing values of the $J_\mathrm{3NN}$, $J_\mathrm{3NN}'$, and $J_\mathrm{3NN}''$ parameters.

\begin{figure}[t!]
\includegraphics[width=\linewidth]{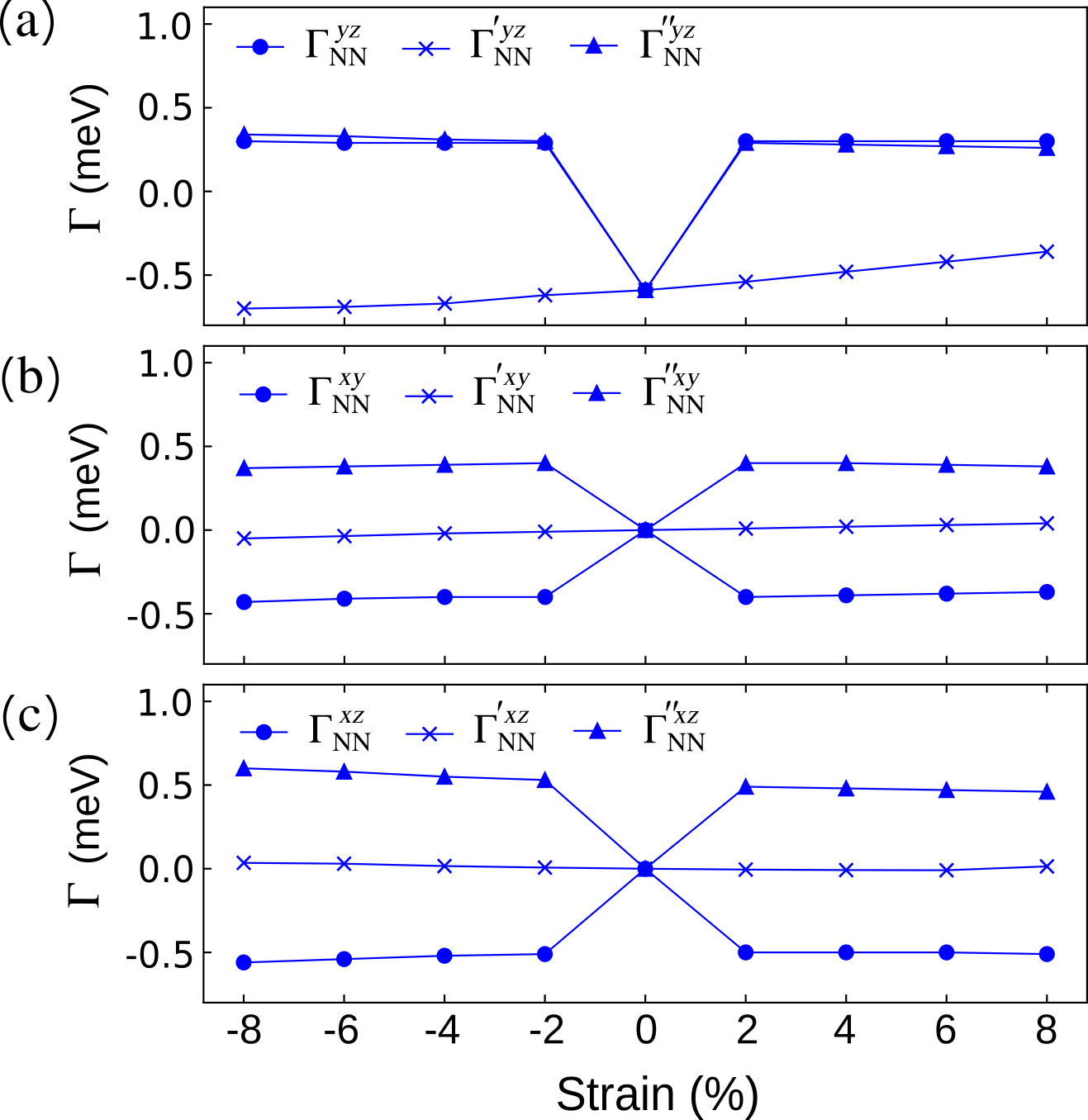}%
\caption{\label{fig:unioffdia} The anisotropic off-diagonal components of exchange tensor (a) \(\Gamma_\mathrm{NN}^{yz}\), (b) \(\Gamma_\mathrm{NN}^{xy}\), and (c) \(\Gamma_\mathrm{NN}^{xz}\), for NiI$_2$ monolayer, as a function of uniaxial strain.}
\end{figure}
\begin{figure}[t!]
\includegraphics[width=\linewidth]{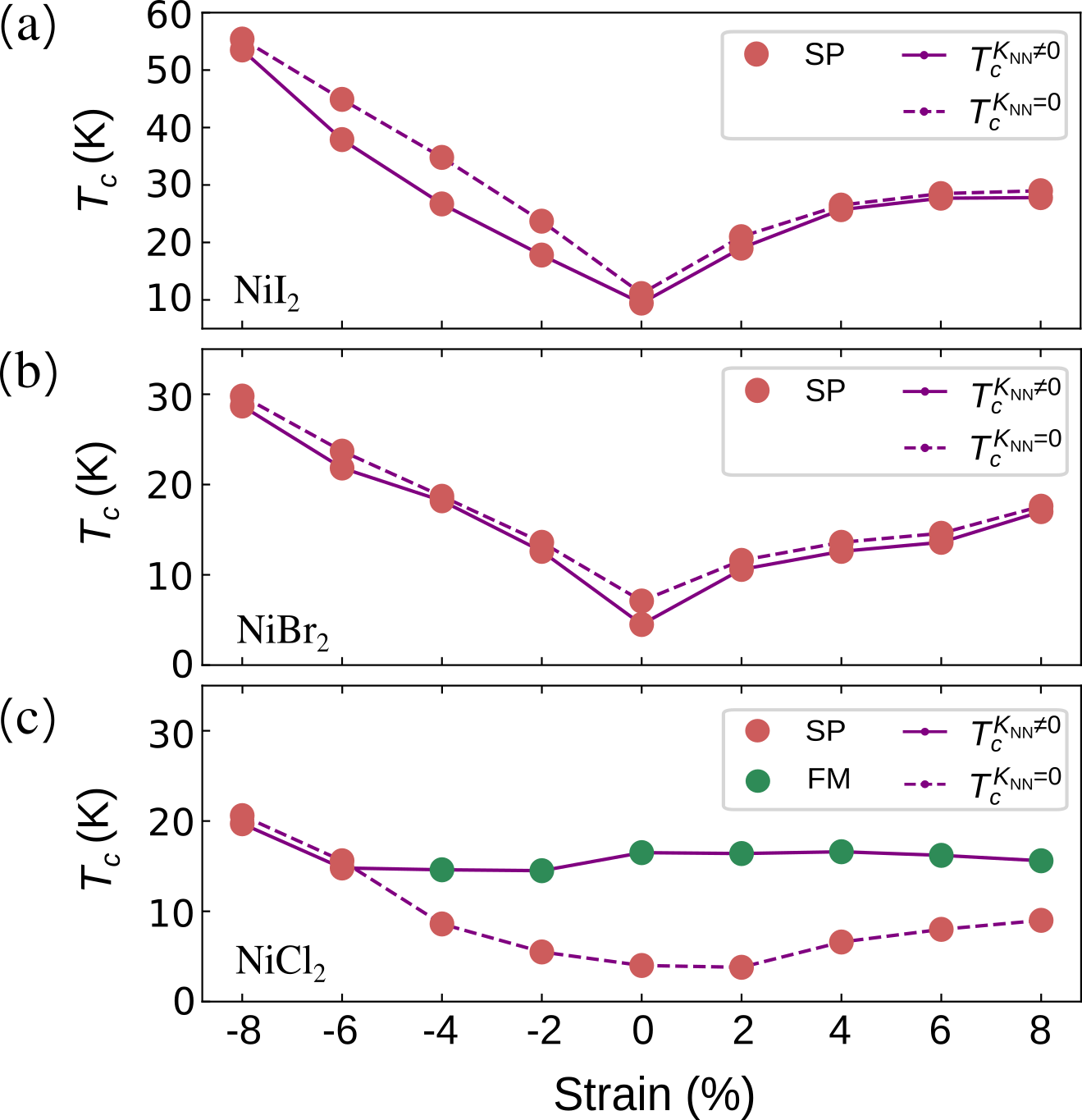}%
\caption{\label{fig:uniaxtc} The critical temperatures as a function of uniaxial strain for monolayers of (a) NiI$_2$, (b) NiBr$_2$, and (c) NiCl$_2$, as computed by parallel tempering Monte Carlo simulations. Solid and dashed lines depict critical temperatures computed from the Heisenberg Hamiltonian, with ($T_c^{K_\mathrm{NN} \neq 0}$) and without ($T_c^{K_\mathrm{NN} = 0}$) consideration of biquadratic exchange interactions. The color-coded dots indicate the ground-state spin order corresponding to each strain: green dots represent ferromagnetic (FM) order, and red dots signify spin-spiral (SP) order.}
\end{figure}

As shown in Figure~\ref{fig:uniaxpa}(d-f), the NN-BQ exchange interaction ($K_\mathrm{NN}$, $K_\mathrm{NN}'$, and $K_\mathrm{NN}''$) remains nearly constant under different uniaxial strain conditions, maintaining values between -0.6 meV and -0.8 meV.

In the context of NN anisotropic exchange parameters, the impact of uniaxial strain on the components of the exchange tensor can be different from that of biaxial strain. 
As shown in Figure~\ref{fig:unioffdia}, the application of uniaxial strain on NiI$_2$ not only induces nonzero values for the \(\Gamma_\mathrm{NN}^{yz}\) component but also for \(\Gamma_\mathrm{NN}^{xy}\) and \(\Gamma_\mathrm{NN}^{xz}\) components. This is in contrast to the effect of biaxial strain, where \(\Gamma_\mathrm{NN}^{xy}\) and \(\Gamma_\mathrm{NN}^{xz}\) were zero. On the other hand, in NiCl$_2$ and NiBr$_2$ monolayers, similar to the behavior observed under biaxial strain, they exhibit isotropic behavior under uniaxial strain. The discontinuities observed in Figures~\ref{fig:uniaxpa} and ~\ref{fig:unioffdia} at 0\% strain arise from the breaking of rotational symmetry. At 0\% strain, the system retains this symmetry, enforcing the equivalency of both the biquadratic and anisotropic exchange parameters across all bonds. However, once uniaxial strain is applied, the symmetry is broken, leading to a variation in these parameters.

For a more comprehensive understanding of such behavior of the exchange tensor parameters, we refer the reader to the Supplementary Material~\cite{supplm}.

As depicted in Figure~\ref{fig:uniaxtc}(a,b), NiI$_2$ and NiBr$_2$ display similar trends in the changes of both critical temperatures ($T_c^{K_\mathrm{NN} \neq 0}$ and $T_c^{K_\mathrm{NN} = 0}$). They maintain the SP as their ground-state spin texture. Notably, when subjected to compressive strain, $T_c^{K_\mathrm{NN} \neq 0}$ increases to 54 K for NiI$_2$ and 29 K for NiBr$_2$. Similarly, under tensile strain, these critical temperatures rise to 28 K for NiI$_2$ and 17 K for NiBr$_2$. 
In the case of NiCl$_2$, the $T_c^{K_\mathrm{NN} \neq 0}$ exhibits an almost linear trend under both compressive and tensile strains, maintaining the ground-state FM order. However, an exception is observed under a compressive strain of 6\%, where the critical temperature increases to a maximum value of 20 K. This increase coincides with a phase transition from ground-state spin texture to the SP order. For $T_c^{K_\mathrm{NN} = 0}$, an upward trend is observed under both compressive and tensile uniaxial strains. The ground state is found to be in SP order under all these strains. These strong differences between $T_c^{K_\mathrm{NN} = 0}$  and $T_c^{K_\mathrm{NN} \neq 0}$ validate once again the initial premise of this work that the BQ exchange interactions are of significant importance in nickel-dihalides.

\subsection{\label{sec:magnons}Magnonic properties}
Applying strain to the NiX$_2$ monolayers will influence their magnonic properties in two different ways: \emph{(i)} strain can result in a phase transition to another magnetic state with significantly different magnonic behavior, and \emph{(ii)} modulation of the exchange parameters due to the application of strain can result in enhancement or depletion of certain features in the magnonic dispersion. In this section, we show that the former occurs for both NiI$_2$ and NiCl$_2$ under biaxial strain, and illustrate the latter for NiBr$_2$ under biaxial and uniaxial strain. 
\begin{figure}[t!]
\includegraphics[width=\linewidth]{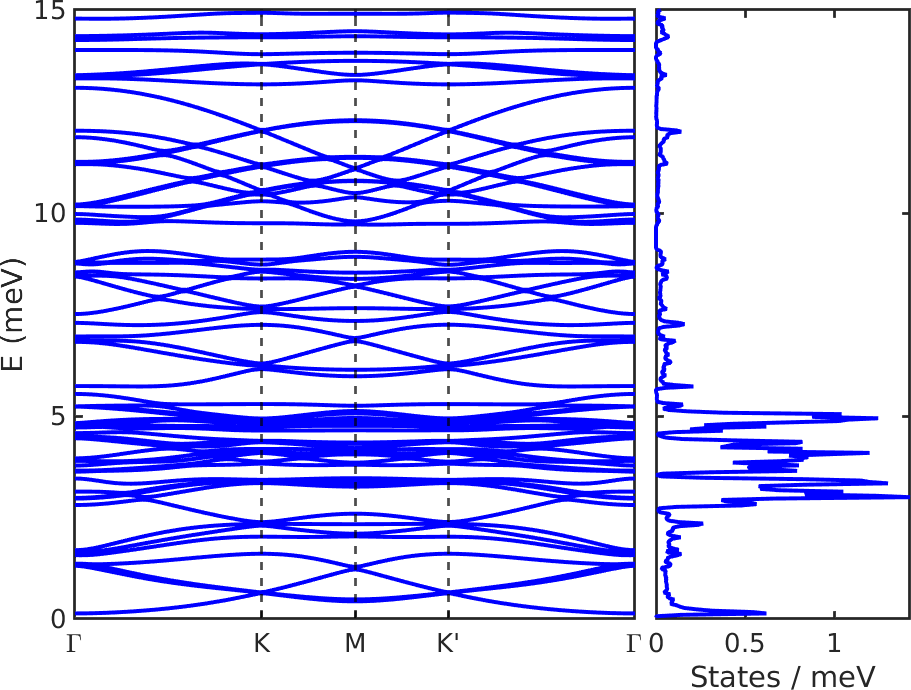}%
\caption{\label{fig:nii2_disp_skx} Magnonic dispersion and DOS of monolayer NiI$_2$ under -4\% biaxial compressive strain in the A2Sk phase.}
\end{figure}

\subsubsection{\label{sec:biaxial_magnons}Effect of biaxial strain}
As shown in Fig. \ref{fig:biaxtc}, application of biaxial strain to monolayer NiI$_2$ can result in a phase transition from the SP to the A2Sk phase for compressive strain with magnitudes greater than -4\%, also resulting in qualitatively very different magnonic properties. In Fig. \ref{fig:nii2_disp_skx}, we show the magnonic dispersion and density of states (DOS) for monolayer NiI$_2$ under -4\% biaxial compressive strain for which the system is characterized by a A2Sk ground state. Notably, there are bandgaps in the dispersion signifying that the A2Sk phase effectively functions as a magnonic crystal since spin waves will not be able to propagate for these frequencies, which is a well known property of skyrmionic lattices \cite{ma2015,chen2021review,wang2020}. The composite Chern number, a quantity used to characterize the topology of degenerate bands \cite{soenen2023stacking,Zhao2020}, equals zero for each group of degenerate bands indicating that the bandgaps are trivial and that no topologically protected edge states are present. In contrast, the magnonic dispersion in the SP and FM phases does not exhibit any bandgaps. 

\begin{figure}[tp!]
\centering
\hspace*{\fill}
\subfloat[\raggedright(a)]{\includegraphics[width=.45\linewidth]{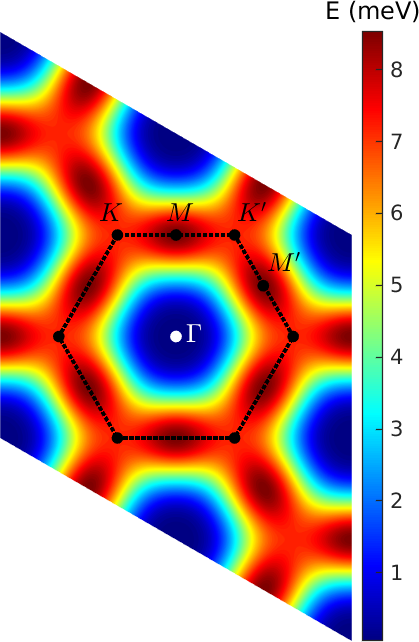}}
\hfill
\subfloat[\raggedright(b)]{\includegraphics[width=.45\linewidth]{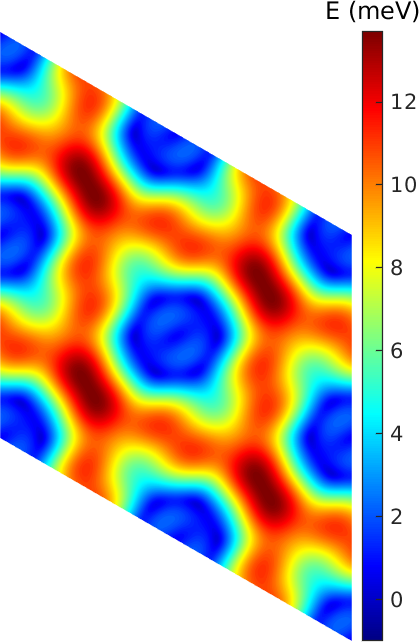}}
\hspace*{\fill}
\caption{\label{fig:magnons1} Magnonic dispersion for (a) monolayer NiCl$_2$ under 4\% biaxial tensile strain which has a FM spin configuration, and (b) monolayer NiBr$_2$ under -6\% biaxial compressive strain which has a SP ground state. The first Brillouin zone and high symmetry points are indicated in (a).}
\end{figure}
\begin{figure}[t!]
\includegraphics[width=\linewidth]{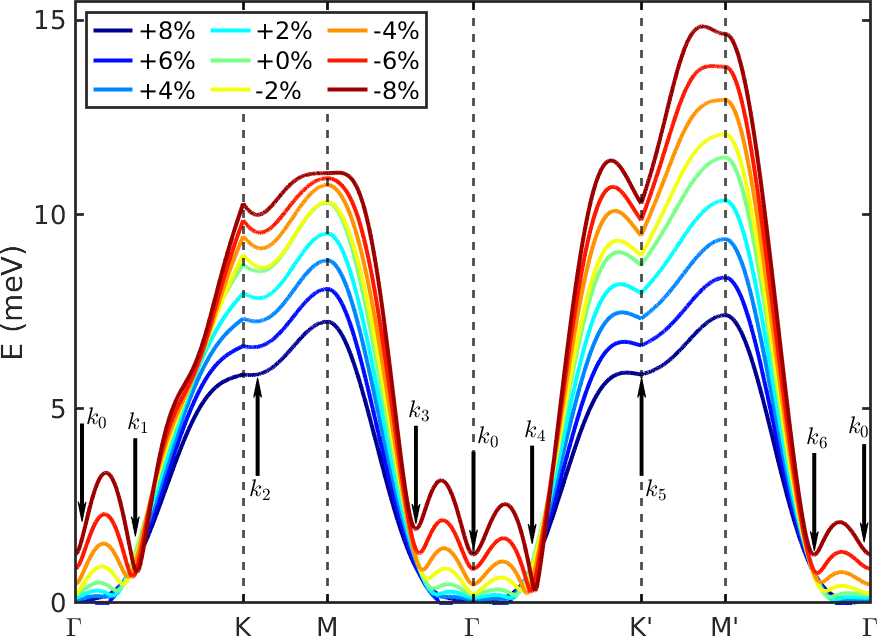}%
\caption{\label{fig:nibr2_disp_biaxial} Effect of biaxial strain on the magnonic dispersion of monolayer NiBr$_2$. For all strain values, the system is characterized by a SP ground state resulting in several (local) minima marked as $k_0 - k_6$, which are modulated by strain.}
\end{figure}

The calculated magnonic dispersion of monolayer NiCl$_2$ under 4\% biaxial tensile strain and monolayer NiBr$_2$ under -6\% biaxial compressive strain are presented in Fig. \ref{fig:magnons1}(a) and Fig. \ref{fig:magnons1}(b), respectively, illustrating the differences in the dispersion between the FM spin state of the former and the SP spin configuration of the latter. Note that the magnetic propagation vector of the SP phase lies parallel to the $\Gamma-K$ path. In the FM phase, there are degeneracies at both the $K$ and $K'$ points and the $M$ and $M'$ points due to the rotational symmetry. However, in the SP spin configuration, the latter symmetry is broken, lifting the degeneracy of the $M$ and $M'$ points. Furthermore, in the SP phase, several local extrema appear in the dispersion, which can be distinguished more easily by plotting the magnon dispersion of strained NiBr$_2$ along the $\Gamma-K-M-\Gamma-K'-M'-\Gamma$ path in reciprocal space, as shown in Fig. \ref{fig:nibr2_disp_biaxial}. The presence of these local minima, recognized as soft magnon modes, are in good agreement with with recent works \cite{cong2024soft,olsen2024,rybakov2024}, who predicted similar modes for NiI$_2$ and NiBr$_2$. Interestingly, in NiCl$_2$, application of tensile strain gives rise to a phase transition between the SP and FM phases, hence biaxial strain serves as a tool to induce or eliminate these soft magnon modes in this material. 

Fig. \ref{fig:nibr2_disp_biaxial} shows how the magnonic dispersion of the SP phase of monolayer NiBr$_2$ changes as a function of strain. First and foremost, notice that in each case we only find finite magnon frequencies, which is attributed to the presence of anisotropy in the system  \cite{menezes2022,soenen2023strain,soenen2023stacking,cong2024soft}. Furthermore, the absence of imaginary frequencies indicates that the SP phase is stable for each of the considered strain values \cite{soenen2023stacking,cong2024soft}. Several interesting points along the $\Gamma-K-M-\Gamma-K'-M'-\Gamma$ path are marked in Fig. \ref{fig:nibr2_disp_biaxial}. Although $k_2$ and $k_5$ appear to be local minima along this specific path in the BZ, they are in fact saddle points and are, thus, excluded from further discussion. The other points are true (local) minima and correspond to the aforementioned soft magnon modes. For FM materials, e.g. strained NiCl$_2$ (see Fig. \ref{fig:magnons1}(a)) or the chromium trihalides \cite{menezes2022,soenen2023strain,soenen2023stacking}, the global energy minimum of the magnonic dispersion typically occurs at the $\Gamma$ point, however, for the SP phase in NiBr$_2$ the mode at $k_4$ has lower energy for all considered strain values. 

Further, note that strain has a clear influence on the magnon dispersion, particularly on the soft magnon modes. Application of compressive strain increases the overall magnon frequency but also significantly sharpens the local minima corresponding to the soft magnon modes. Just like the SP phase itself, the soft magnon modes are stabilized with increasing $|J_\mathrm{3NN}/J_\mathrm{NN}|$ ratio. The strain-dependent behavior for the SP phases in NiI$_2$ and NiCl$_2$ resembles that of NiBr$_2$, hence the same conclusions can be drawn for those materials.   

\subsubsection{\label{sec:uniaxial_magnons}Effect of uniaxial strain}
\begin{figure}[t!]
\includegraphics[width=\linewidth]{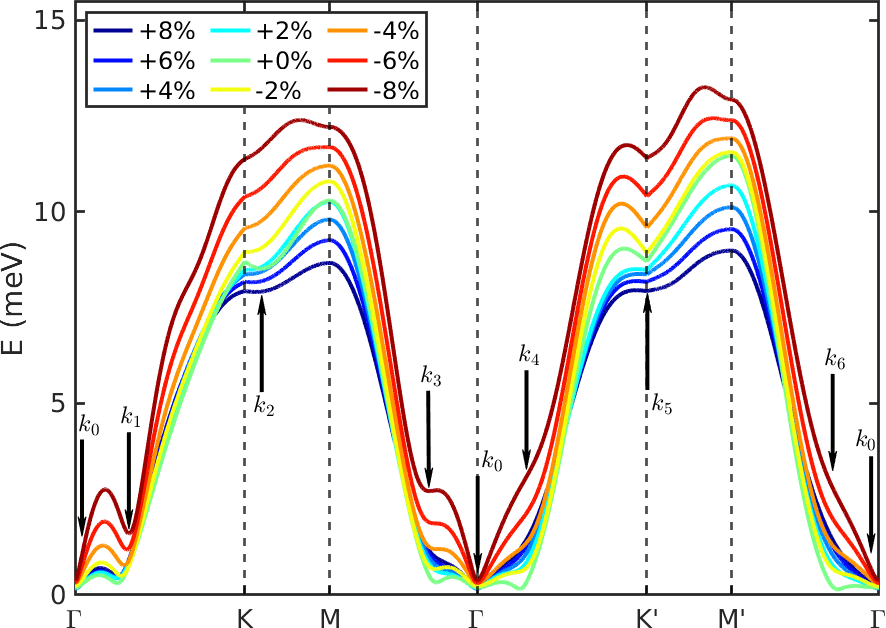}%
\caption{\label{fig:nibr2_disp_uniaxial} Effect of uniaxial strain on the magnonic dispersion of monolayer NiBr$_2$. For all strain values, the system is characterized by a SP ground state resulting in several (local) minima marked as $k_0 - k_6$, which are modulated by strain.}
\end{figure}

As shown in Fig. \ref{fig:uniaxtc}, application of uniaxial strain does not result in magnetic phase transitions when the BQ exchange is not included in the Heisenberg Hamiltonian. For all considered strain values, all three NiX$_2$ (X = I, Br, Cl) monolayers have a SP ground state. In Fig. \ref{fig:nibr2_disp_uniaxial}, we illustrate the effect of uniaxial strain tuning on the magnonic dispersion of NiBr$_2$ only, being representative for all three compounds. In contrast to the biaxially strained case, application of tensile strain does not result in softening of the magnon modes. Interestingly, for both the case of compressive and tensile strain, the (local) minima at the $k_4$ and $k_6$ points vanish, meanwhile the minima at the $k_1$ and $k_3$ points shift up in energy, with as a result that the mode at $k_0$ is located in the global minimum of the dispersion. This result is in agreement with earlier work \cite{cong2024soft} that describes how the soft magnon modes will diminish and eventually be depleted upon inclusion of the Kitaev interaction in their model. Application of uniaxial strain, both compressive and tensile strain, results in anisotropic tuning of the exchange couplings (see Fig. \ref{fig:uniaxpa}), which is exactly how the Kitaev interaction is defined in the model of Cong et al. \cite{cong2024soft}. In contrast, application of biaxial strain tunes does not increase the Kitaev interaction and hence preserves the soft magnon modes in the dispersion.   

\section{\label{sec:conclusion}conclusion}

In summary, using a combination of DFT and MC simulations we explored the magnetic properties of the 2D multiferroic monolayer NiI$_2$ and other compounds from the same family, namely NiBr$_2$ and NiCl$_2$, under lateral strain. When unstrained, we found that NiI$_2$ and NiBr$_2$ exhibit spin-spiral (SP) ground states, in agreement with previous experimental observations \cite{amini2024atomic,2021noncollinear}. Similarly, we validated that NiCl$_2$ exhibits a ferromagnetic (FM) ground-state order, only after we included the nearest-neighbor biquadratic (NN-BQ) exchange coupling in the Hamiltonian. Furthermore, our study demonstrates that the inclusion of NN-BQ exchange leads to different critical temperatures in comparison with models that do not include this interaction. With that in mind, we proceeded to identify both biaxial and uniaxial strain as potent tools for finely tuning the magnetic order in NiX$_2$ (X = I, Br, Cl) monolayers. By applying strain, one can deliberately tailor the degree of magnetic frustration in these systems, leading to a diverse magnetic phase diagram that includes SP, FM, and skyrmionic (antibiskyrmion, A2Sk) phases for NiI$_2$ and SP and FM states for both NiBr$_2$ and NiCl$_2$. Additionally, strain drastically alters the critical temperature of the corresponding phases. Notably, for NiI$_2$, the critical temperature increases from 11.6 K in the pristine case to values up to 37.0 K and 54.0 K for an 8\% compressive strain in the biaxial and uniaxial cases, respectively. We here reemphasize the significance of considering the biquadratic exchange interaction in understanding the magnetic properties of the NiX$_2$ monolayers and underscore its role in stabilizing different magnetic states, as well as the magnonic properties that are very sensitive to the magnetic state and its response to strain. Overall, our work establishes strain engineering as an effective avenue for tailoring magnetic and magnonic properties in NiX$_2$ monolayers, as well as their transition temperatures, thereby expanding the applicability of these materials in emergent spintronic devices based on 2D magnetic materials and heterostructures.

\section*{ACKNOWLEDGMENTS}
We thank Denis \v{S}abani for useful discussions. This work was supported by the Research Foundation-Flanders (FWO-Vlaanderen). The computational resources used in this work were provided by the VSC (Flemish Supercomputer Center), funded by Research Foundation-Flanders (FWO) and the Flemish Government -- department EWI.

\bibliography{aapmsamp}

\end{document}